\documentclass[prl,twocolumn,showpacs,preprintnumbers,amsmath,amssymb,tightenlines,epsfig,superscriptaddress]{revtex4}
\usepackage{graphicx}
\usepackage{color}
\usepackage[percent]{overpic}
\newcommand{\bra}[1]{\langle\,{#1}\, |}
\newcommand{\ket}[1]{|\,{#1}\,\rangle}

\newcommand{\ssection}[1]{{\noi  \it #1:}}

\newcommand{\sub}[2]{{#1}_{\mbox{\!\! \scriptsize #2}}}

\newcommand{\bv}[1]{\mathbf{ #1 }}

\def\noi{\noindent}
\def\beq{\begin{equation}}
\def\eeq{\end{equation}}

\def\CR{\nonumber\\[0.15cm]}

\newcommand{\rref}[1]{Ref.~\cite{#1}}
\newcommand{\fref}[1]{Fig.~\ref{#1}}
\newcommand{\frefp}[2]{Fig.~\ref{#1}~(#2)}

\newcommand{\eref}[1]{Eq.~(\ref{#1})}

\newcommand{\cref}[1]{chapter~\ref{#1}}

\newcommand{\Cref}[1]{Chapter~\ref{#1}}

\newcommand{\bref}[1]{(\ref{#1})}

\usepackage{ulem}  
\begin{document}

\title{Phase-Imprinting of Bose-Einstein Condensates with Rydberg Impurities}
\author{Rick~Mukherjee}
\affiliation{Max Planck Institute for the Physics of Complex Systems, N\"othnitzer Strasse 38, 01187 Dresden, Germany}
\author{Cenap~Ates}
\affiliation{School of Physics and Astronomy, University of Nottingham, Nottingham, NG7 2RD, UK}
\author{Weibin~Li}
\affiliation{School of Physics and Astronomy, University of Nottingham, Nottingham, NG7 2RD, UK}
\author{Sebastian~W{\"u}ster}
\affiliation{Max Planck Institute for the Physics of Complex Systems, N\"othnitzer Strasse 38, 01187 Dresden, Germany}
\email{sew654@pks.mpg.de}
\begin{abstract}
We show how the phase profile of Bose-Einstein condensates can be engineered through its interaction with localized Rydberg excitations.
The interaction is made controllable and long-range by off-resonantly coupling the condensate to another Rydberg state with laser light. Our technique allows the mapping of entanglement generated in systems of few strongly interacting Rydberg atoms onto much larger atom clouds in hybrid setups. As an example we discuss the creation of a spatial mesoscopic superposition state from a bright soliton. Additionally, the phase imprinted onto the condensate using the Rydberg excitations is a diagnostic tool for the latter. For example a condensate time-of-flight image would permit reconstructing the pattern of an embedded Rydberg crystal.
\end{abstract}

\maketitle

\ssection{Introduction} 
The imprinting of tailored phase profiles onto the complex order parameter of a Bose-Einstein condensate (BEC) \cite{dobrek:phaseimp} is a versatile tool for the creation of topological states such as 
solitons \cite{denschlag:phaseimp,burger:solimprint} and vortices \cite{dobrek:phaseimp} or even Skyrmions \cite{ruoste:imprint1}. More generally imprinting allows the transfer of the BEC into a desired state of atom flow or motion. Typically the phase is generated purely optical using lasers. We propose to engineer phases exploiting the interactions between BEC atoms that are 
Rydberg dressed~\cite{santos:dressing,nils:supersolids,pupillo:dressed,honer:dressing,fabian:bullets,wuester:dressing,johnson:dressing,balewski:dressingNJP,jau:expt} and resonantly excited Rydberg atoms \cite{heidemann:rydexcBEC,viteau:rydexclattice,balewski:elecBEC,gaj:molspecdensshift,karpiuk:imagingelectrons}. This phase imprinting will rely 
on the matter-wave coherence of the condensate, and thus represent an instance of genuine Rydberg-BEC physics. 

We show that phase imprinting creates a versatile interface between ultra-cold Rydberg- and BEC physics. Firstly, it allows entangled Rydberg states \cite{jaksch:dipoleblockade,rick:Rydberglattice,wuester:cannon} to be mapped onto the many-body wave function of the condensate. As one example we discuss how to turn an atomic Bell state $\ket{+}=(\ket{Rg} +\ket{gR})/\sqrt{2}$ (with two atomic electronic states $\ket{g}$, $\ket{R}$) into a spatial mesoscopic superposition state in the position of a single BEC bright soliton, akin to the proposal of \cite{weiss:solitoncat}.
Secondly, phase imprinting represents a tool to probe Rydberg electronic states via their effect on a condensate \cite{middelkamp:rydinBEC,karpiuk:imagingelectrons,olmos:amplification,guenter:EIT,guenter:EITexpt,wang:rydelecBEC}. To demonstrate, we show
signatures of Rydberg-crystals \cite{pohl:crystal} in expected condensate time of flight spectra.

Our scheme relies on weakly admixing Rydberg character to all the atoms in a condensate cloud through far off-resonant laser coupling between their stable ground state and a highly excited Rydberg state \cite{nils:supersolids,pupillo:dressed}, as experimentally demonstrated for two atoms \cite{jau:expt}. All these "dressed" atoms then interact with some previously prepared, fully excited atoms in a different Rydberg state, {referred to as impurity or control atom depending on whether they are inside or outside the condensate. The atomic species of impurities may be identical to condensate atoms. }

{The interaction involving impurities} can be made dominant over the simultaneously induced long-range condensate self-interaction \cite{nils:supersolids,pupillo:dressed}. {Using this we} demonstrate the imprinting of sizeable phases for realistic parameters.
\begin{figure}[htb]
\includegraphics[width=0.99\columnwidth]{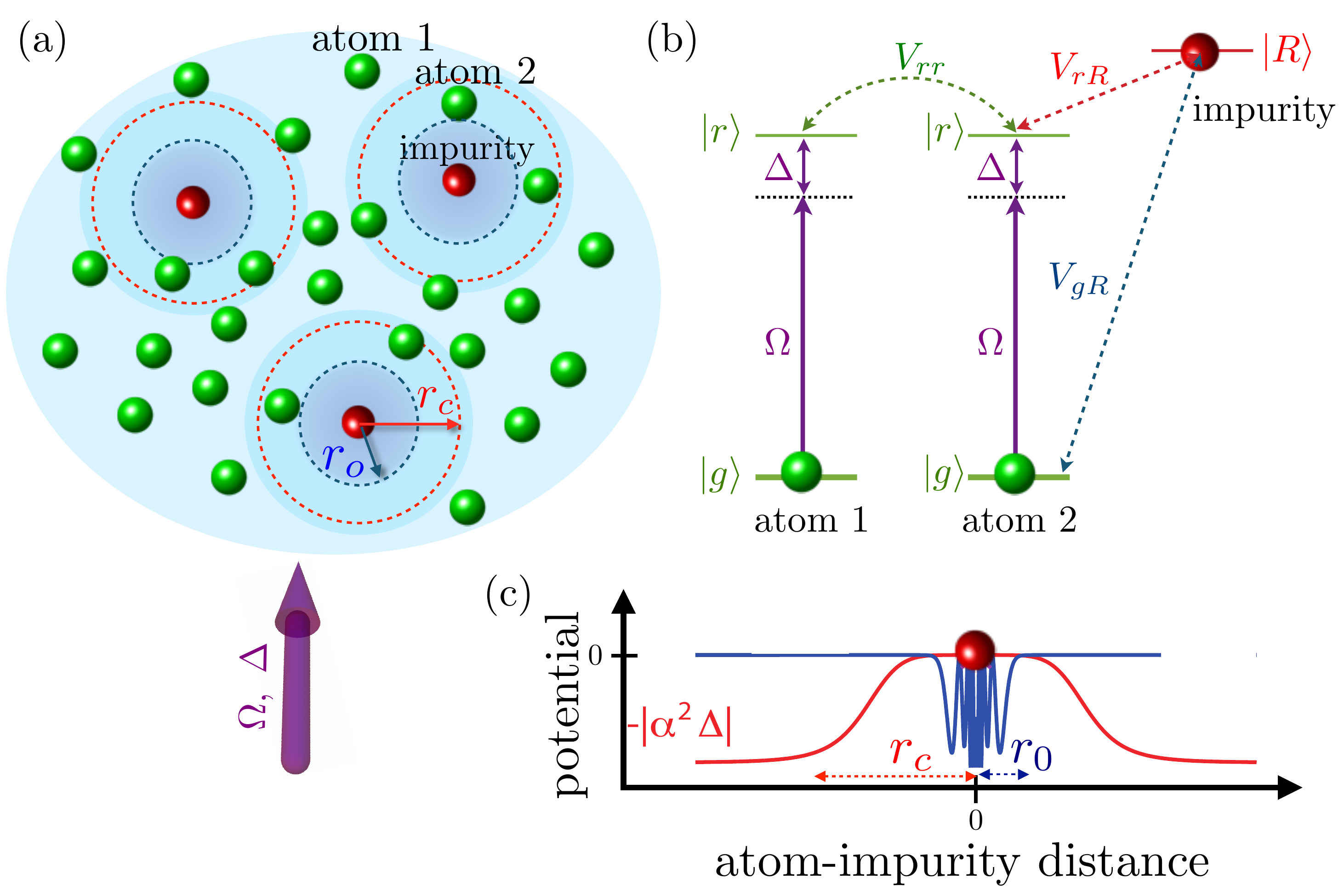}
\caption{\label{sketch}(color online) (a) The setup. Several Rydberg excited impurities (red balls) are embedded in a Bose-Einstein condensate. The condensed atoms (green balls) mainly interact with impurities through dressing induced long-range interactions of characteristic range $r_c$ (red dashed circles). At distances~$d$ within the Rydberg orbit, $d<r_0$, collisions with Rydberg electrons are also relevant (blue dashed circles). After an imprinting period, the condensate phase (blue shades) will be modified only in the vicinity of impurities (dark blue).
(b) Level scheme for three representative atoms. Interactions $V_{rr}$ give rise to long-range interactions between dressed condensate atoms as in \cite{nils:supersolids,pupillo:dressed}. Interactions $V_{rR}$ and $V_{gR}$ give rise to stronger potentials between dressed condensate atoms and impurities, which are the main focus here. (c) Sketch of these potentials $\sub{U}{eff}^{(2)}$ (red, see text) and $V_{gR}$ (blue) near one of the impurities. $\alpha=\Omega/(2\Delta)$ quantifies the degree of Rydberg admixture.
}
\end{figure}

The condensate phase is also affected by direct collisions of the Rydberg electron with condensate atoms in the ground state, without dressing coupling. This can be exploited to visualise the electron orbital through condensate densities \cite{karpiuk:imagingelectrons,wang:rydelecBEC}. In contrast the imprinting through dressed interactions discussed here extends the spatial scale, smoothness and controllability of phase profiles. The use of Rydberg impurities and effectively one- or two dimensional (1D,2D) condensates in this article circumvents {some} of the interaction strength and many-body related problems of dressing discussed in~\cite{balewski:dressingNJP}.

\ssection{Interactions between Rydberg impurities and dressed atoms} 
Consider a gas of $N$ {Rb} atoms with mass $M$ at locations $\bv{R}_n$, as depicted in \frefp{sketch}{a}. They may be in either of three electronic states: ground state $\ket{g}$, Rydberg state $\ket{r}=\ket{\nu s}$ or Rydberg state $\ket{R}=\ket{\nu' s}$, with principal quantum numbers $\nu$, $\nu'$, where $\nu <\nu'$ as shown in \frefp{sketch}{b}. The label $s$ implies angular momentum $l=0$.  We write the system Hamiltonian $\sub{\hat{H}}{tot}= \sub{\hat{H}}{0} + \sub{\hat{H}}{dress} + \sub{\hat{H}}{int}$, using the notation $\hat{\sigma}^{(n)}_{kk'}=\ket{k}\bra{k'}$, where $k$, $k'$ $\in \{g,r,R\}$ and $\hat{\sigma}^{(n)}_{kk'}$ acts on atom $n$ only:
\begin{align}
 \sub{\hat{H}}{0} &=\sum_{n=1}^{N}\big[ -\hbar^2 \nabla^2_{\bv{R}_n}/(2M)  +  W(\bv{R}_n) \hat{\sigma}^{(n)}_{gg}  \big],
 \CR
\sub{\hat{H}}{dress} &=\sum_{n=1}^{N}\big[\Omega(t)\hat{\sigma}^{(n)}_{rg}/2   + \mbox{h.c.} -\Delta\hat{\sigma}^{(n)}_{rr}  \big],
\CR
\sub{\hat{H}}{int}& =\sum_{a,b\in \{g,r,R\}}\sum_{n\neq m=1}^{N}   V_{ab}(d_{nm})\hat{\sigma}^{(n)}_{aa}\hat{\sigma}^{(m)}_{bb}.
\label{Hamiltonian}
\end{align}
Ground state atoms experience an external trapping potential $W(\bv{R}_n)$. The state $\ket{g}$ is coherently coupled to  $\ket{r}$ with Rabi frequency $\Omega(t)$ and detuning $\Delta$. Defining {$d_{nm} = |\bv{R}_n-\bv{R}_m|$}, we take van-der-Waals interactions between two Rydberg atoms in states $a$, $b$ as $V_{ab}(d)=C^{(ab)}_6/[(1+\delta_{ab})d^6]$ for simplicity, where $C^{(ab)}_6$ is the dispersion coefficient and $\delta_{ab}$ Kronecker's delta. Between ground state atoms we assume the usual contact interaction $V_{gg}(d)=g\delta(d)$, where $g=4\pi \hbar^2 a_s/M$ with atom-atom s-wave scattering length $a_s$. Finally, ground state and Rydberg electrons interact via Fermi pseudo potentials $V_{g,r/R}(d)=V_0 |\Psi_{\nu/\nu'}(d)|^2$ with
$V_0=2\pi \hbar^2 a_e/m_e$~\cite{greene:ultralongrangemol}, electron mass $m_e$, electron-atom scattering length $a_e$~\cite{footnote:momdep} and Rydberg orbital wave function $\Psi_{\nu/\nu'}(d)$. 
Our examples will be based on Rb with $a_e=-0.849\,$nm, assuming Rydberg states $|r\rangle=\ket{55S}$ and $\ket{R}=\ket{76S}$ \cite{footnote:interactions}.

Now consider a scenario where $\sub{N}{imp}$ of the $N$ atoms have been excited to the impurity Rydberg state $\ket{R}$, denoting their locations by $\{ \bv{x}_n\}\subset\{\bv{R}_n \}$. Many different random or deterministic location patterns can be created, depending on the method of excitation. 
We do not consider the excitation step, but refer to the literature on selective optical access \cite{nogrette:hologarrays,ScTiKr11_907_,rvb:patternedexcitation}, the exploitation of blockade effects \cite{pohl:crystal,gaerttner:floatcrystals,weimer:phases,wuester:echo,lesanovsky:nonequil_structures,cenap:emergent,rvb:crystalchirped,schempp:rydaggstatistics,schauss:mesocryst,schauss:dyncryst} or on condensate density dependent energy shifts \cite{balewski:elecBEC,middelkamp:rydinBEC}. Interactions $V_{RR}$ between two Rydberg impurities are important in the stage of impurity placement but can subsequently be neglected in the examples discussed here. The light-atom coupling in $\sub{\hat{H}}{dress}$ will cause long-range interactions for all atoms, which would otherwise be present only among Rydberg excited atoms ($\ket{r}$, $\ket{R}$). Assuming far off-resonant coupling between $\ket{g}$ and $\ket{r}$, so that $|\alpha| \ll 1$ for $\alpha = \Omega(t)/(2\Delta)$, we determine these interactions using fourth order perturbation theory in $\Omega(t)$. {Calculating} the energy shift $\sub{\Delta E}{$\bv{g}\bv{R}$}$ of the state $\ket{0}=\ket{\bv{g}\bv{R}}$ where the $\sub{N}{imp}$ atoms at locations $\bv{x}_n$ are in $\ket{R}$ and the other $\bar{N}=N-\sub{N}{imp}$ atoms in $\ket{g}$, we obtain $\sub{\Delta E}{$\bv{g}\bv{R}$}=\alpha^2 E_{(2)}+ \alpha^4 E_{(4)}$, where $E_{(2)}=\Delta \sum_n^{\bar{N}} (1 - \sum_m^{\sub{N}{imp}} V_{rR}(|\bv{R}_{n} -\bv{x}_m|)/\Delta )^{-1}$. We find $E_{(4)}$ in comparison negligible, see \cite{sup:info}.

In this article, impurities will only affect dynamics for very short times, such that their motion can be assumed frozen in space \cite{anderson1998}, and they also do not undergo state changes.
Similar to \cite{nils:supersolids} we merge the effective interactions obtained through the laser dressing with the direct interactions between atoms contained in \bref{Hamiltonian} ($\sub{V}{gR}$, $\sub{V}{gg}$) to arrive at the following effective Gross-Pitaevskii equation (GPE) for the dressed and condensed ground-state atoms in the presence of Rydberg impurities: 
\begin{align}
&i\hbar \frac{\partial}{\partial t}\phi(\bv{R})= \bigg(-\frac{\hbar^2}{2 m}\nabla^2 + W(\bv{R}) + g|\phi(\bv{R})|^2 
\CR
&+ \bigg[ \sub{U}{eff}^{(2)}(\bv{R},\{\bv{x}_m\},t) +\sum_m^{\sub{N}{imp}} V_0 |\Psi(|\bv{R} - \bv{x}_m|)|^2 \bigg] \bigg)\phi(\bv{R}),
\CR
&\sub{U}{eff}^{(2)}=\alpha^2(t) \Delta  (1 - \sum_m^{\sub{N}{imp}} V_{rR}(|\bv{R} -\bv{x}_m|)/\Delta )^{-1}.
\label{dressedGPE}
\end{align}

Here the presence of a few impurity atoms that are fully in a Rydberg state causes strong, long-range interactions with the remaining atoms, which can be treated as  external single body potential for the condensate. The corresponding terms in \bref{dressedGPE} are in square brackets and will be the central tool of the present work.  Note that the dominant part $E_{(2)}$ is orders of magnitude larger than the dressed interaction between condensate atoms, causing quite different physics than the latter \cite{nils:supersolids}. The induced potentials are sketched in \frefp{sketch}{c} as red line ($\sub{U}{eff}^{(2)}$, using signs $V_{rR}/\Delta<0$) and blue line ($|\Psi(|\bv{R}|)|^2$) for a single impurity. Either potential is associated with an important length scale. The plateau of the dressing induced potential extends to the critical radius $r_c=(C_6^{(rR)}/|\Delta|)^{1/6}$, which also sets the width of the region of potential drop. The extent of the direct interaction potential $V_{gR}$ is the radius of the Rydberg electron orbital $\Psi_{\nu'}$ of the impurity, $r_0\approx a_0\nu'^2$ with $a_0$ the Bohr radius. We focus on parameters for which molecular resonances are avoided and also $r_c >  r_0$ \cite{Bendkowsky:ultralongmol,Bendkowsky:boundquantumreflection,Li:homonuc}. Although included in our solutions of \eref{dressedGPE}~\cite{footnote:cutoff}, the direct interactions $V_{gR}$ then play a minor role. It has been shown that many-body perturbative calculations as used here are valid only as long as 
there would be much less than one Rydberg excitation blockade sphere ($\sub{N}{bl}\ll1$) \cite{footnote:nbl,honer:dressing}, which will be satisfied here.
 
Our applications of \eref{dressedGPE} to phase-imprinting involve two dynamical stages: In a first short stage of duration $\sub{\tau}{imp}\sim 10\mu$s, the condensate order parameter $\phi$ acquires a dynamical phase $\varphi(\bv{R})$, such that $\phi(\bv{R})\rightarrow \exp{[i \varphi(\bv{R})]}\phi(\bv{R})$ with $\varphi(\bv{R}) =- \sub{U}{eff}^{(2)}(\bv{R},\{ \bv{x}_m\}) \sub{\tau}{imp}$. The time $\sub{\tau}{imp}$ and strength of $\sub{U}{eff}^{(2)}$ are such that other energies can be neglected. Only in this stage are dressed interactions enabled though $\sub{\hat{H}}{dress}$. In a much longer second stage ($t\sim 10$ms), the condensate evolves according to the usual GPE, and the initially imprinted phase profile is typically transformed into condensate flow and/or density variations. We consider two examples that highlight the main strengths of Rydberg phase imprinting: Transferring entanglement from a Rydberg system onto a BEC, and inferring the geometry of a collection of Rydberg impurities in a cold gas.

\ssection{Entanglement transfer} 
{We first consider a 1D arrangement of a  $^{85}$Rb BEC bright soliton (see \rref{book:solitons,wuester:collsoll} and references therein) with $\bar{N}=400$ atoms, located between two individual atoms, which are each tightly trapped in their own optical tweezer at $x_{1,2}$ with position spread $\sub{\sigma}{con}=0.05\,\mu$m. The atoms outside the condensate are referred to as control atoms. As shown in \fref{solimprint}, the control atoms are separated by a distance $D=|x_1-x_2|=3\,\mu$m. The soliton requires attractive contact interactions, $g<0$,
which can be achieved using the Feshbach resonance~\cite{mahir:review} at $B\sim155$ G in $^{85}$Rb \cite{jila:nova}, we assume $a_s(B)=-5.33$ nm \cite{footnote:oneDreduction}.
Only the control atoms  are now driven into the Rydberg state $\ket{R}=\ket{\nu'}$ under blockade conditions, resulting in an entangled two-body state $\ket{+}=(\ket{gR} + \ket{Rg})/\sqrt{2}$. Subsequently we enable the dressing coupling $\Omega(t)$ to the state $\ket{r}$ for the bulk soliton, resulting in the dressed potential sketched blue in \frefp{solimprint}{a,b}, which depends on whether the left or right control atom was originally excited. After an adiabatically enabled and disabled imprinting period $\sub{\tau}{imp}=36\, \mu$s, with $\Omega(t)/h=3$ MHz, $\Delta/h=-500$ MHz, the condensate has acquired the phase profile shown in red. Following de-excitation of the control atoms, we allow $\sub{\tau}{mov}=2$ ms of free evolution according to the first line of \eref{dressedGPE}, i.e.~$\Omega(t)=0$. After $\sub{\tau}{mov}$, the soliton has moved by about $2\, \mu$m to the left or right depending on the imprinted phase profile, as shown in \frefp{solimprint}{c,d}. Let us denote the many-body wave function of the gas for these two cases as $\sub{\ket{\Psi}}{left/right}$. The entire process should be quantum coherent, since the initial imprinting happens well before a control Rydberg state or dressed Rydberg state from the condensate would decay \cite{footnote:lifetime}, resulting in a final many-body state $\ket{\sub{\Psi}{+}}=(\sub{\ket{\Psi}}{left} + \sub{\ket{\Psi}}{right})/\sqrt{2}$.

The tightly trapped control atoms \cite{Li:kuzmich:atomlighentangle} and the small length scales \cite{nogrette:hologarrays} are technical challenges for the above proposal. However, when these are overcome, one obtains a mesoscopic entangled state \cite{weiss:solitoncat,hallwood:robustflow,gordon:BECcat,dunningham:BECarrays,moebius:bobbels,cirac:BECsuperpos,ng:mesoscop}, where the entire soliton of $\bar{N}$ atoms is in a superposition of two different locations.  As proposed in \cite{weiss:solitoncat,moebius:cat}, 

The superposition nature of the resulting state can be proven interferometrically \cite{weiss:solitoncat,moebius:cat,gertjerenken:quantclassscatt:pra} upon recombination, for which one would additionally place the soliton into a weak harmonic trap $W(\bv{R})$. De-coherence processes during the creation of such a highly entangled many-body state limit $\bar{N}$, but are small for our choice here \cite{footnote:atomloss}.  A full quantum many-body treatment including coherence between control atom states and condensate atoms may be subject of further research. 
%
%
%
%
\begin{figure}[htb]
\begin{overpic}[width=0.99\columnwidth]{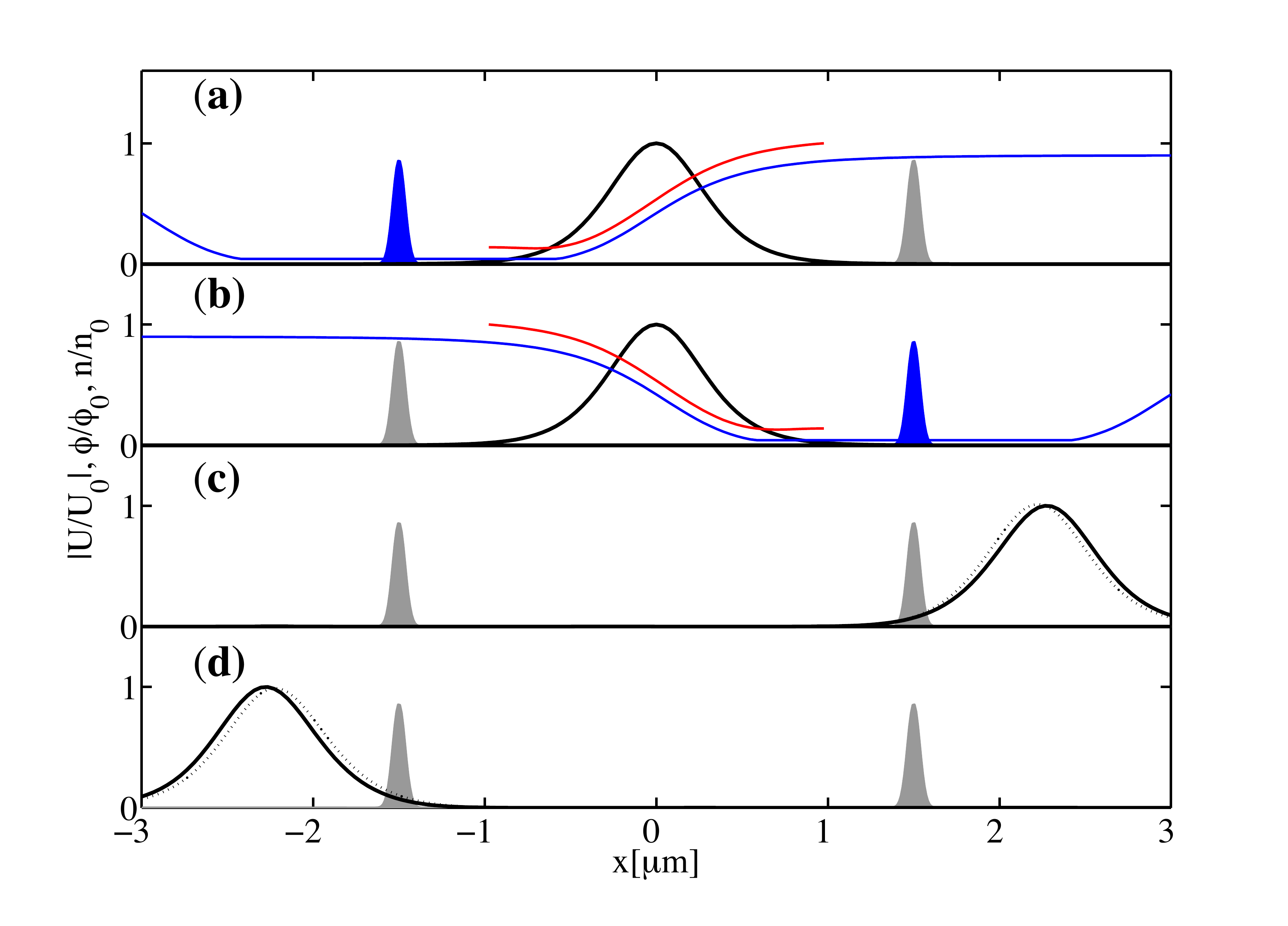}
 \put (24,62) {\large$x_1\downarrow$}
 \put (63.5,62) {\large$x_2\downarrow$}
  \put (75.5,24.5) {\footnotesize$\sub{\ket{\Psi}}{right}$}
   \put (15.4,10.5) {\footnotesize$\sub{\ket{\Psi}}{left}$}
 \end{overpic}
 \caption{\label{solimprint}(color online) Entanglement transfer from two control Rydberg atoms onto a mesoscopic BEC cloud. The position space density of control atoms is shown as shaded curves, blue for state $\ket{R}$ and gray for state $\ket{g}$. The condensate density for a soliton is shown as thick black line. (blue) Dressing potential $|\sub{U}{eff}^{(2)}|$, (red) condensate phase $\varphi$ after imprinting. We show all quantities with arbitrary normalization to fit the same axis. (a,b) Initial state at $t=\sub{\tau}{imp}$, (c,d) final state at $t=\sub{\tau}{imp}+\sub{\tau}{mov}$. (a,c) show the control atom configuration $\ket{Rg}$, (b,d) show $\ket{gR}$. For each configuration, we model condensate evolution separately using \eref{dressedGPE} \cite{footnote:oneDreduction}. The dotted line in (c,d) is for shifted control atom positions $x_i +  2 \sub{\sigma}{con}$. }
\end{figure}
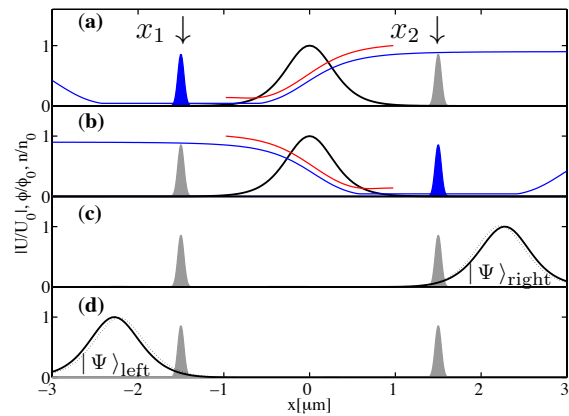
%

\ssection{Rydberg crystal imprinting}
The maximally entangled state $\ket{\sub{\varphi}{+}}$ is but one example of entanglement arising due to strong Rydberg-Rydberg interactions. Another example is given by
 spatially ordered (crystal) structures formed by large numbers of Rydberg excitations in a cold gas \cite{schauss:mesocryst,pohl:crystal,weimer:crystal,rvb:crystalchirped,schachenmeyer:latticecrystal,gaerttner:crystal}. We show now that Rydberg-phase-imprinting in the presence of such structures leads to condensate momentum spectra that allow the reconstruction of the locations of impurities. 

Let us consider a 2D model \cite{footnote:quasi2D} of a BEC confined in a pancake shaped harmonic trap $W(\bv{R})=m[\omega_r^2 (x^2 +y^2) + \omega_ z^2 z^2]/2$, with frequencies $\omega_z\gg\omega_r$ and $a_s=5.5\,$nm, thus $g>0$.
 Using a scheme as discussed in \cite{pohl:crystal}, $\sub{N}{imp}$ impurities can be arranged, for example, in a crystal like structure within the condensate cloud.

 \begin{figure}[htb]
 	\includegraphics[width=0.99\columnwidth]{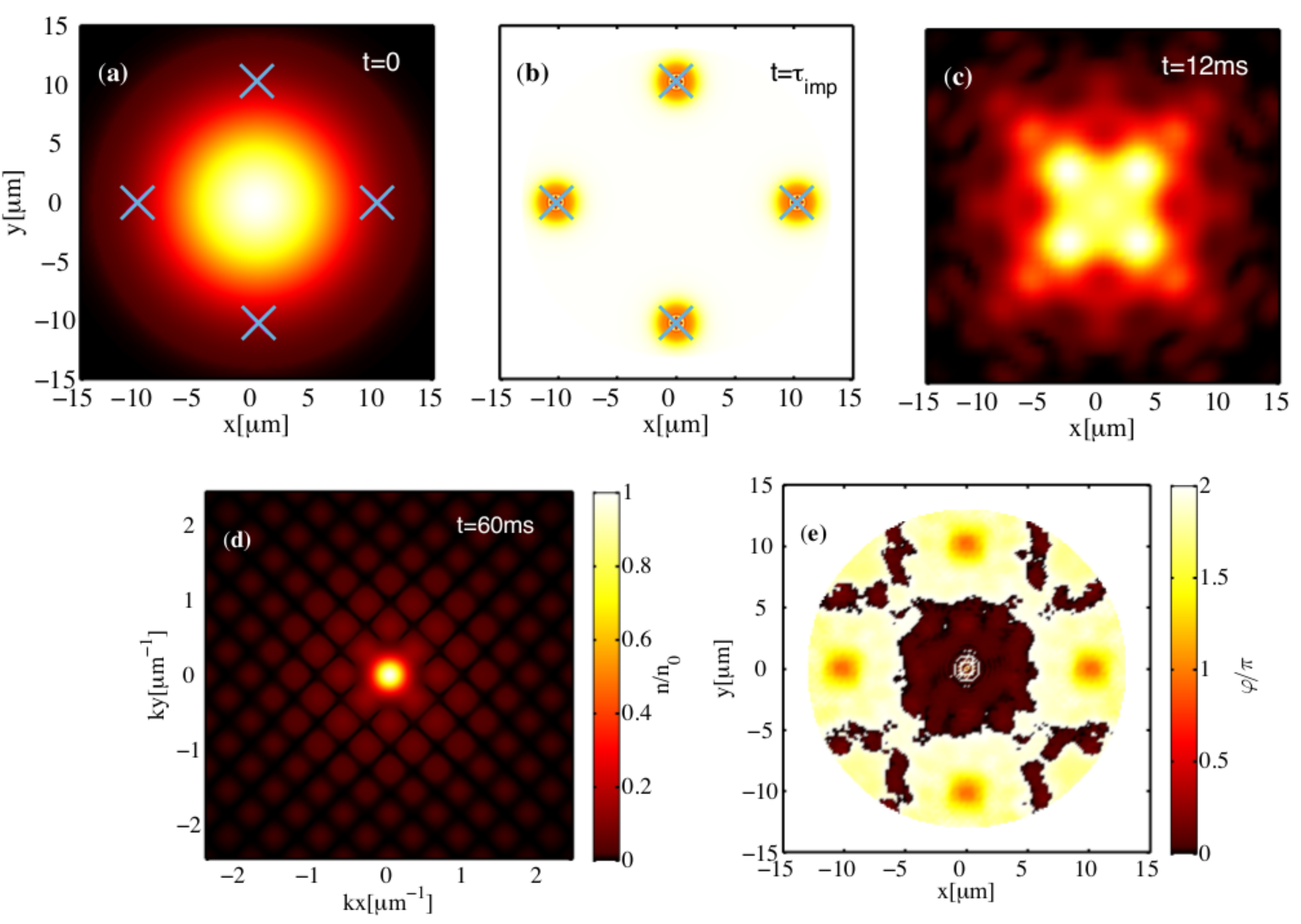}
 	\caption{\label{clover}(color online) Determining the spatial arrangement of Rydberg impurities (shown as blue crosses) via phase imprinting in a 2D $^{87}$Rb BEC of $N=100$ atoms in a trap with $\omega_r=(2\pi)\,2$Hz and $\omega_z=(2\pi)\,100$Hz. Colorbar for (a,c,d) see (d), where $n_0$ is the respective peak density, colorbar for (b,e), see (e). (a) Initial condensate density $\rho=|\phi(\bv{R})|^2$. (b) Condensate phase $\varphi$ following phase imprinting ($t=\sub{\tau}{imp}$). (c) Condensate density shortly after imprinting ($t=\sub{\tau}{imp} + 12$ms).  (d) Final momentum spectrum $|\tilde{\phi}(k)|$  ($t=60$ ms),  (e) Phase profile reconstructed from (d) as described in the text.}
 \end{figure}

 For a distribution of impurities as shown in \frefp{clover}{a}, we numerically solve \eref{dressedGPE} \cite{xmds:paper,xmds:docu} enabling the potentials $\sub{U}{eff}^{(2)}$ and $V_{gR}$ \cite{footnote:cutoff} for a short imprinting period $\sub{\tau}{imp}=18.5\,\mu$s only, using $\Omega/h =4$ MHz, $\Delta/h =-150$ MHz. This is followed by evolution under the influence of the contact interactions, {but with} disabled harmonic trap and the impurities assumed removed via field-ionisation \cite{book:gallagher,balewski:elecBEC}. At some final time where momentum spectra no longer significantly change, we plot the expected time-of-flight (TOF) images in \fref{clover}. We also show the position space density shortly after phase-imprinting. 

If the effect of atomic collisions, described by the non-linear term $g|\phi(\bv{R})|^2$, is not too large, the final momentum spectrum is roughly the Fourier transform of $\phi(\bv{R})=\sqrt{\sub{n}{ini}(\bv{R})}\exp[i\varphi(\bv{R})]$, where $\sub{n}{ini}(\bv{R})$ is the known initial atom density in the trap, and $\varphi(\bv{R})$ the phase profile generated through imprinting and shown in \frefp{clover}{b}. A standard phase-retrieval algorithm \cite{gerchberg_saxton,fienup:comparison} is then able to recover the phase profile as shown in \frefp{clover}{e} from which impurity positions can clearly be inferred. 
The algorithm relies on iterative Fourier-transforms involving two known quantities: the final time-of-flight image from which the
\emph{modulus} of the condensate order parameter is extracted $|\tilde{\phi}(k)|$, and the initial condensate density $\sub{n}{ini}(\bv{R})$. We briefly describe it in \cite{sup:info}.

We find that simple phase retrieval fails for the case of \fref{clover} but larger condensate densities due to condensate self-interactions. This might be remedied by more sophisticated variants of the phase retrieval algorithm \cite{fienup:comparison}, or modifications of atomic interactions using a Feshbach resonance~\cite{mahir:review}. One could then use larger atom clouds, in which case impurity locations can be recovered from a single image as in \frefp{clover}{d} without the need for image alignment in an ensemble average, as in~\cite{schauss:mesocryst}.

The setup just discussed complements Rydberg crystal detection based on single-atom addressing~\cite{schauss:mesocryst} or electromagnetically induced transparency
~\cite{olmos:amplification,guenter:EIT,guenter:EITexpt} by working with a bulk gas and moving the signal from the light to the atomic density. Beyond crystal detection, it enables  
phase-profiles that are otherwise difficult to achieve, for example those akin to \frefp{clover}{b} arising from a crystalline impurity distribution on the surface of a 3D sphere.

\ssection{Conclusions and outlook} 
{We proposed a novel phase imprinting technique for Bose-Einstein condensates, employing long-range interactions between}{ condensate atoms and embedded Rydberg excited impurity atoms,  
created}{ by coupling condensate atoms far off-resonantly to another Rydberg state.} The scheme offers functionalities beyond existing imprinting methods, as it allows mapping of entanglement from few-body Rydberg states onto the whole atom cloud, and strengthens BEC as a diagnostic tool for {detecting} Rydberg excitations in an ensemble of atoms. We illustrate the former through a proposal for the creation of a mesoscopic entangled state in the position of a cloud of atoms and the latter by exploring the link between Rydberg crystal structures in a condensate, and momentum space spectra after phase-imprinting. 

Combing the techniques discussed here with imprinting effects by a Rydberg electron in a larger orbital
may offer additional possibilites, due to the unusual shape of the Rydberg orbital \cite{balewski:elecBEC,gaj:molspecdensshift,karpiuk:imagingelectrons}.
Other interesting physics might arise from the interplay of phase-imprinting and controlled impurity motion \cite{cenap:motion,moebius:cradle,wuester:cradle,wuester:CI,leonhardt:switch}.

\acknowledgments
We gladly acknowledge fruitful discussions with Igor Lesanovsky, Thomas Pohl, Rick van Bijnen and Shannon Whitlock, as well as 
EU financial support received from the Marie Curie Initial Training Network (ITN) ÔCOHERENCE". W. L. is supported through the Nottingham Research Fellowship by the University of Nottingham.

\appendix
\section{Supplement: Phase-Imprinting of Bose-Einstein Condensates with Rydberg Impurities}

\ssection{Dressed ground state -- impurity interaction}
%
Here we analyze effective interactions arising from $V_{rR}$ and $V_{rr}$ in (1) of the main article perturbatively. Consider $\hat{H}'=\sum_{n=1}^{N}[\Omega(t)\hat{\sigma}^{(n)}_{gr}   + \mbox{h.c.}]/2$ the perturbation and $\sub{\hat{H}}{0}=\sub{\hat{H}}{tot}-\hat{H}'$ the unperturbed Hamiltonian. We define states
\begin{align}
|0\rangle &= |g_1 \ g_2 \hdots g_{\bar{N}} R_1 \hdots R_{\sub{N}{imp}} \rangle \ , \CR
|i\rangle  &= |g_1 \ g_2 \hdots r_i \hdots g_{\bar{N}} R_1 \hdots R_{\sub{N}{imp}} \rangle \ ,\CR
|ij \rangle  &= |g_1 \ g_2 \hdots r_i \hdots r_j \hdots g_{\bar{N}} R_1 \hdots R_{\sub{N}{imp}} \rangle \ ,
\end{align} 
with zero-, one- and two atoms in the Rydberg state $\ket{r}$, and a fixed number of $\sub{N}{imp}$ impurities in the Rydberg state $\ket{R}$. 

Let us introduce the shorthand $U^{(ab)}_{nl}=V_{ab}(|\mathbf{R}_n - \mathbf{R}_l|)$. We then calculate corrections $\sub{E}{0}$ to the energy of the state $\ket{0}$. Up to fourth order perturbation theory in $\hat{H}'$ we obtain $\sub{E}{0}=E_{(0)} +  \alpha^2 E_{(2)}+ \alpha^4 E_{(4)}$, where we define the unperturbed energy $E_{(0)}=0$. We find to leading order
\begin{align}
\alpha^2  E_{(2)}= \alpha^2\Delta \sum_{n=1}^{\bar{N}} \frac{1}{1 -\sum_{m=1}^{\sub{N}{imp}} U^{(rR)}_{nm}/\Delta } .
\label{background_impurity2}
\end{align}
Since the dressing parameter $\alpha=\Omega/(2\Delta)\ll1$, this will be the dominant consequence of laser dressing.
The fourth order correction is
\begin{widetext}
\begin{align}
\!\!\!\!\!\!\!\!\!\!\!\!\!\!\!\!\!\!\!\!\!\!\!\!\!\!\!\!\!
\alpha^4 E_{(4)}=& \left(\frac{\Omega}{2}\right)^4 \bigg(  \sum^{\bar{N}}_{n,n'; n\neq n'} \bigg[ \frac{1}{(\Delta - \sum_{m=1}^{\sub{N}{imp}} U^{rR}_{nm})(\Delta - \sum_{m=1}^{\sub{N}{imp}} U^{rR}_{n'm})(2\Delta - \sum_{m=1}^{\sub{N}{imp}} (U^{rR}_{nm} + U^{rR}_{n'm}) - U^{rr}_{nn'})}\CR
&+ \frac{1}{(\Delta - \sum_{m=1}^{\sub{N}{imp}} U^{rR}_{nm})^2(2\Delta - \sum_{m=1}^{\sub{N}{imp}} (U^{rR}_{nm} + U^{rR}_{n'm}) - U^{rr}_{nn'})} \bigg]
- \sum^{\bar{N}}_{n,n'}\frac{1}{(\Delta - \sum_{m=1}^{\sub{N}{imp}} U^{rR}_{nm})(\Delta - \sum_{m=1}^{\sub{N}{imp}} U^{rR}_{n'm})^2} \bigg) .
\label{background_impurity4}
\end{align}
\end{widetext}

To understand the physics contained in $E_{(4)}$ and ultimately find a simple approximation, we analyse $E_{(4)}$ in some limiting cases.
When all interactions vanish:
\begin{align}
\lim_{U^{(ab)}_{nl} \rightarrow 0}  \alpha^4 E_{(4)} =- \Delta\alpha^4  \bar{N} ,
\end{align}
as expected for the fourth order light shift of $\bar{N}$ non-interacting background atoms.

If we let all background atoms interact, but remove the effect of impurity atoms (e.g. placing them far away), we obtain
\begin{align}
\lim_{U^{(rR)}_{nl} \rightarrow 0}   \alpha^4 E_{(4)} =- \Delta\alpha^4  \bar{N}  
+ \Delta\alpha^4  \sum^{\bar{N}}_{n\neq n'}\frac{U^{rr}_{nn'}}{2\Delta  - U^{rr}_{nn'}}.
\label{background_background}
\end{align}
This is simply the interaction between two Rydberg-dressed ground state atoms that is known from \cite{nils:supersolids}.

If instead we neglect interactions between background atoms the expression becomes
\begin{align}
\lim_{U^{(rr)}_{nl} \rightarrow 0}   \alpha^4 E_{(4)} = -  \Delta \alpha^4 \sum_n \frac{1}{(1 - \sum_m^{\sub{N}{imp}} U^{(rR)}_{nm}/\Delta )^3}.
\label{background_impurity4}
\end{align}
This is the fourth order contribution to the effective interaction of one background atom and the impurity atom, the leading contribution of which is $\alpha^2 E_{(2)}$.

Beyond these limits, \bref{background_impurity4} must contain nontrivial three-body physics, as becomes clear when we consider the impact of an impurity atom on the effective interaction between two background atoms, \bref{background_background}: Whenever \emph{either} of two atoms $1$ or $2$ separately falls into the blockade radius of the impurity, it cannot reach its Rydberg state $\ket{r}$, and the interaction \bref{background_background} must be suppressed. 

Here we only wish to consider the dominant consequences of the induced interactions. The interaction \bref{background_background} is
suppressed by $\alpha^2$ compared to \bref{background_impurity2}, however due to the much larger number of background atoms compared to impurity atoms, care has to be taken when estimating the importance of the term.

In order to simplify this estimate and (for the sake of completeness) allow an inclusion of the effects of \bref{background_impurity4} in a Gross-Pitaevskii description,
we propose the following simplification of $E_{(4)}$, based on the discussion above
\begin{widetext}
\begin{align}
 \alpha^4 \tilde{E}_{(4)} &=\sum_n^{\bar{N}} \frac{- \Delta \alpha^4 }{\left( 1-\sum_m^{\sub{N}{imp}} U^{(rR)}_{nm}/\Delta \right)^3}  + \Delta \alpha^4 \sum^{\bar{N}}_{n,n'; n'\neq n}\frac{U^{rr}_{nn'}}{2\Delta  - U^{rr}_{nn'}}
\left[  \frac{1}{\left( 1-\sum_m^{\sub{N}{imp}} U^{(rR)}_{nm}/\Delta \right)^3} \right].
\label{approxE4}   
\end{align}
\end{widetext}
The term in square brackets is a phenomenological screening factor that suppresses effective background-background interactions whenever atom $n$ is too close to an impurity (we see shortly why we do not include atom $n'$ in the screening factor). It can be easily verified that \bref{approxE4} agrees with \bref{background_impurity4} in the limiting cases discussed earlier. Since these span most of parameter space, the approximation of the original \bref{background_impurity4} works usually well. We consider deviations for a special case with two background atoms and one impurity atom. 
Deviations are expected for geometries where the distances of all three atoms are comparable and of the order of the blockade radius. This is confirmed in \fref{expression_test}.
Deviations are small enough for our purpose, which is merely to ascertain that fourth order interactions \bref{approxE4} can be neglected for parameters here.
\begin{figure}[htb]
\includegraphics[width=0.99\columnwidth]{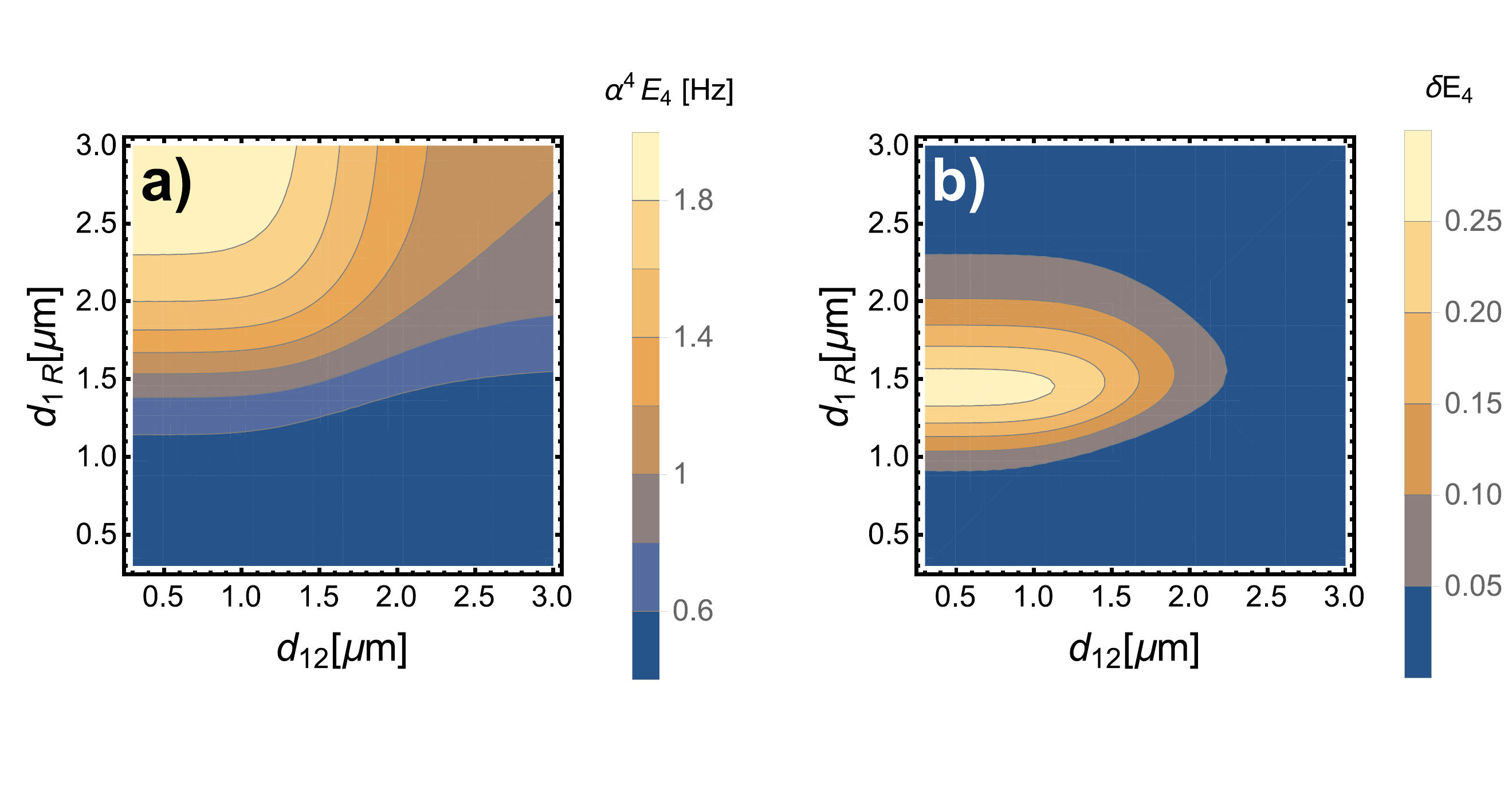}
\caption{\label{expression_test}(color online) Benchmark of our simplification for the fourth order dressed potential $ \alpha^4 \tilde{E}_{(4)} $ \bref{approxE4}. (a) Full potential $ \alpha^4 E_{(4)} $ \bref{background_impurity4} for fixed $d_{2R}=4.7\mu$m. (b) Relative difference $\delta E=|\tilde{E}_{(4)}-E_{(4)}|/E_{(4)}$. 
\label{approxbenchmark}}
\end{figure}

In fact we only employed \eref{approxE4} to confirm that for all cases discussed in detail in this article, $\alpha^4 E_{(4)}$ can be neglected relative to $\alpha^2 E_{(2)}$. The continuous space version of \bref{approxE4} would give rise to a term
\begin{align}
&\sub{U}{eff}^{(4a)}(\bv{R}, \{\bv{x}_m\},t)  \phi(\bv{R})\CR
+&\int d^3\bv{R}'  \sub{U}{eff}^{(4b)}(\bv{R},\bv{R}', \{\bv{x}_m\},t)  |\phi(\bv{R'})|^2 \phi(\bv{R})
\label{GPterm}
\end{align}
with
\begin{align}
&\sub{U}{eff}^{(4a)}(\bv{R}, \{\bv{x}_m\} ) = - \Delta \alpha^4 ( 1-\sum_m^{\sub{N}{imp}} U^{(rR)}(\bv{R} - \bv{x}_m)/\Delta )^{-3}, \CR
 &\sub{U}{eff}^{(4b)}(\bv{R},\bv{R'}, \{\bv{x}_m\} )=\sub{U}{eff}^{(rr)}(\bv{R}-\bv{R'})\sub{f}{scr}(\bv{R}, \{\bv{x}_m\} ),\CR
 &\sub{U}{eff}^{(rr)}(\bv{R}-\bv{R'}) = \Delta\alpha^4  \frac{U^{rr}(\bv{R}-\bv{R'})}{2\Delta  - U^{rr}(\bv{R}-\bv{R'})}, \CR
 &\sub{f}{scr}(\bv{R}, \{\bv{x}_m\} )= (1-\sum_m^{\sub{N}{imp}} U^{(rR)}(\bv{R}-\bv{x}_m) / \Delta)^{-3} ,
\label{Ueff4_continuous}
\end{align}
on the rhs of the GPE (2) of the main article. Here it becomes clear why the approximation involved in \eref{approxE4} is advantageous: We can now write the non-local interactions in \eref{GPterm} as
\begin{align}
&\int d^3\bv{R}'  \sub{U}{eff}^{(4b)}(\bv{R},\bv{R}', \{\bv{x}_m\},t)  |\phi(\bv{R'})|^2 \phi(\bv{R})=\CR
&\left[ \int d^3\bv{R}' [\sub{U}{eff}^{(rr)}(\bv{R}-\bv{R'})  |\phi(\bv{R'})|^2 \right]\sub{f}{scr}(\bv{R}, \{\bv{x}_m\} ) \phi(\bv{R}),
\label{convol_simp}
\end{align}
where the term in square brackets can be efficiently evaluated via a convolution as usual. Had the screening factor in \bref{approxE4} contained $n'$ this would not have been possible. 

For all results in the main article, an interaction \eref{convol_simp} would give rise to a dynamical phase of $\varphi<0.003\pi$ during $\sub{\tau}{imp}$.

\ssection{Phase recovery algorithm by Gerchberg and Saxton}
%
We briefly summarize here the algorithm due to Gerchberg and Saxton \cite{gerchberg_saxton} for the recovery of phase information of a function from just the \emph{intensity} (density) of the Fourier transform of the function. The algorithm and variants \cite{fienup:comparison} are widely applied, e.g.~in x-ray crystallography.

Consider a pair of function $\phi(\bv{R})$ and its Fourier transform $\tilde{\phi}(\bv{k})$ defined by:
\begin{align}
\tilde{\phi}(\bv{k}) &= {\cal F}[\phi(\bv{R})] = (2\pi)^{-3/2} \int d^3 \bv{R} e^{-i \bv{k} \bv{R}} \phi(\bv{R}).
\end{align}
For our application here, $\phi(\bv{R})$ and $\tilde{\phi}(\bv{k})$ are the condensate wave function in the position- and momentum representation (respectively), but the algorithm described is more general. Let us define the modulus and phase of a function $f= \rho[f] \exp{(i \varphi{[f]}]})$ with $\rho,\varphi \in \mathbb{R}$.

The problem is to recover $\varphi[\phi]$ when only $\rho[\phi]$ and $\rho[\tilde{\phi}]$ are known. This is solved by repeated application of the following steps~\cite{fienup:comparison} 
\begin{align}
f_k&={\cal F}^{-1}[g_k], \\
f'_k&= \rho[\phi] \exp{(i \varphi[f_k])},
\label{posconst}
\\
g'_k&={\cal F}[f'_k],\\
g_{k+1}&=\rho[\tilde{\phi}] \exp{(i \varphi[g'_k])},
\label{momconst}
\end{align}
where $k$ is the iteration number and the algorithm is initialized with $g_0 = \rho[\tilde{\phi}]$. In \eref{posconst} and \eref{momconst} we are enforcing the known position and Fourier-space density constraints. The algorithm usually converges but possibly very slowly, hence improvements exist \cite{fienup:comparison}. For our purposes (Fig.~3 of the main article), between $2\times 10^3$ and $2\times 10^4$ iterations of the basic algorithm were sufficient.

\bibliography{phase_imprinting_rydberg}

\begin{thebibliography}{81}
\bibitem[{foo({\natexlab{a}})}]{footnote:momdep}
\bibinfo{note}{The momentum dependence of which we neglect here.}
\bibitem[{foo({\natexlab{c}})}]{footnote:cutoff}
\bibinfo{note}{To keep numerics tractable, we include $V_{g,R}(r)$ only
  approximately: (i) Its amplitude, substantially exceeding other energy
  scales, is cut off at $|V_{g,R}(r)|=5 \Delta \alpha^2$. (ii) Its spatial
  profile is undersampled, since numerical discretisations of space that can
  represent the time-of-flight expansion required for Fig.~3 cannot
  simultaneously resolve the very fine oscillations of the Rydberg wave
  function. We have separately confirmed that the relative strength of
  amplitude and phase perturbations of the condensate wavefunction after the
  imprinting period are representative.}
\bibitem[{foo({\natexlab{d}})}]{footnote:nbl}
\bibinfo{note}{In a homogeneous 3D system we have $\sub{N}{bl}=\alpha^2 n_0
  \sub{V}{bl}$ atoms per blockade sphere $\sub{V}{bl}=4\pi r_{bl}^3/3$ (the
  volume in which one Rydberg excitation blocks another), where $n_0$ is the
  background atom density and $r_{bl} = (C_6^{(rr)}/\Delta)^{1/6}$ the blockade
  radius.}
\bibitem[{foo({\natexlab{e}})}]{footnote:oneDreduction}
\bibinfo{note}{For an effective 1D treatment, we replace $g\rightarrow g_{1D}=g
  (2\pi\sigma_\perp^2)^{-1}$, $\sigma_\perp=\sqrt{\hbar/m/\omega_\perp}$,
  $\omega_\perp=(2\pi)150$ Hz.}
\bibitem[{foo({\natexlab{f}})}]{footnote:lifetime}
\bibinfo{note}{The life time of $\ket{R}$ is $ \sub{\tau}{$\nu'=76$} \approx
  180\, \mu$s \cite{beterov:BBR} and the effective dressed condensate life time
  is $\sub{\tau}{eff}=\sub{\tau}{$\nu=55$}/(\bar{N} \alpha^{2})\approx 23$ ms.}
\bibitem[{foo({\natexlab{g}})}]{footnote:atomloss}
\bibinfo{note}{E.g.~we expect a mean loss of $0.8$ atoms from the soliton
  during $\sub{\tau}{mov}$, dominated by three-body recombination.}
\end{thebibliography}
\end{document}